\documentclass{JINST}
\usepackage{graphicx}
\usepackage{caption}
\usepackage{subcaption}
\usepackage{amsmath}

\usepackage[square, comma, numbers, sort&compress]{natbib} 
\title{Cryogenic Gaseous Photomultiplier for Position Reconstruction of Liquid Argon Scintillation Light}
\author{B. Lopez Paredes$^a$\thanks{Corresponding
author.}~, C.D.R. Azevedo$^b$, S. Paganis$^c$\thanks{Most of the work performed
while at University of Sheffield}~, A.L.M. Silva$^b$, N.J.C. Spooner$^a$ and J.F.C.A. Veloso$^b$\\
\llap{$^a$}Department of Physics and Astronomy, University of Sheffield,\\
  Hicks Building, Hounsfield Rd, S3 7RH, United Kingdom\\
\llap{$^b$}I3N, Department of Physics, University of Aveiro,\\
  3810-193, Aveiro, Portugal\\
\llap{$^c$}Department of Physics, National Taiwan University,\\
No 1, Sec 4, Roosevelt Road, Taipei 10617, Taiwan\\
  E-mail: \email{brais.lopez.paredes@cern.ch}}

\abstract{
   Presented here are first tests of a Gaseous Photomultiplier based on a cascade
of Thick GEM structures intended for gamma-ray position reconstruction in liquid argon.
The detector has a MgF$_2$ window, transparent to VUV light,
and a CsI photocathode deposited on the first THGEM.
A gain of $8\cdot10^{5}$ per photoelectron
and $\sim100\%$ photoelectron collection efficiency 
are measured at stable operation settings. 
The excellent position resolution capabilities 
of the detector (better than 100$~\mu$m) at 100~kHz readout rate, 
is demonstrated at room temperature. Structural integrity tests of the detector and seals are successfully performed 
at cryogenic temperatures by immersing the detector in liquid Nitrogen,
laying a good foundation for future operation tests in noble liquids.
}


\begin{document}

\section{Introduction}
Noble liquids (xenon, krypton and argon) are growing in importance in particle physics experiments~\cite{LArExp1, LArExp2, LArExp3, LXeExp1, LXeExp2} and they have been proposed as an alternative to crystals in medical particle detectors such as Positron Emission Tomography (PET) scanners~\cite{LXePETResults}. In PET scanners, the position resolution is limited by the granularity of the $360^{\circ}$ crystal and photomultiplier tube (PMT) arrays, and the energy resolution by the crystal scintillation light yield (NaI(Tl): $38$~ph/keV, BGO: $15\%$ relative to NaI(Tl), GSO: $30\%$, LSO: $75\%$~\cite{bushberg2002essential}) and the photodetector resolution. Compared to conventional scintillation crystals, noble liquids have similar or superior scintillation light yield (liquid argon: $\lesssim51$~ph/keV~\cite{LARPHOTONSPERKEV}, liquid xenon: $\sim40$~ph/keV~\cite{PhysRevC.84.045805}), leading to an improved energy resolution. Furthermore, they are transparent to their own scintillation light and, unlike solid 
state detectors, 
degradation of the medium can be counteracted by continuously circulating the liquid through a purifier. Liquid xenon is commonly used due to its high density and scintillation light wavelength at $178$~nm, offering a better stopping power than the other liquids and the possibility of detecting the light with cryogenic photomultipliers~\cite{HamamatsuR8778}. At lower light wavelengths, such as the peak of argon scintillation ($\sim127$~nm), wavelength shifters must be used to continue working with PMTs~\cite{TPBforLAr}. However, liquid argon is the ideal medium due to its very low cost, although alternative light read-out methods are required to avoid the granularity limitations imposed by PMTs and the efficiency loss introduced by wavelength shifters.

Position sensitive Gaseous Photomultipliers (GPMs) can be manufactured with large active areas and with photocathodes sensitive to UV noble liquid scintillation light, offering a cheap alternative to vacuum and solid state photon detectors and with a position resolution on the order of $100$~$\mu$m. Hole-type micropatterned structures like Thick Gaseous Electron Multipliers (THGEMs) are indispensable components in such GPMs. When arranged in a cascade, with the first structure coated with a thin film of photosensitive material and operated at high voltages immersed in a noble gas, they focus the photoelectrons into the holes and provide additional electrons and positive ions via collisions with the gas atoms. The cascaded structure allows for lower individual operating voltages and discharge probability while increasing the detector gain. Caesium Iodide (CsI) can be used to form a reflective photocathode~\cite{PositionVUV}, with sensitivity to UV light below the $220$~nm threshold with a quantum efficiency 
from $\sim15\%$ for liquid xenon scintillation ($178$~nm) to $>60\%$ for liquid argon scintillation light ($127$~nm) (see Figures~7 and~8 in~\cite{CsIQE}) and a time resolution $<10$~ns~\cite{DallaTorre2011, Breskin2009, Chechik2008}.

The stability of position sensitive GPMs at cryogenic temperatures down to $88$~K
has been tested with positive results~\cite{CryogenicGPMtests}, confirming 
the expected reduction in photoelectron extraction efficiency with increased 
gas density at low temperatures. Liquid xenon scintillation light detection has also been performed
with a GPM detector~\cite{CryogenicGPMtests2011}. In this article the construction, operation and testing of a prototype GPM intended for liquid argon scintillation light is presented. Voltage settings are optimised at room temperature to maximise the gain, and the position resolution is studied. Further tests towards the operation of the detector submerged in liquid argon are also carried out: structural tests in liquid Nitrogen, room temperature multiple-photon position reconstruction and gain stability.

\section{GPM Detector Design and Operation Principle}

\begin{figure}[htbp]
  \begin{center}
    \begin{subfigure}{0.35\textwidth}
      \includegraphics[width=\textwidth]{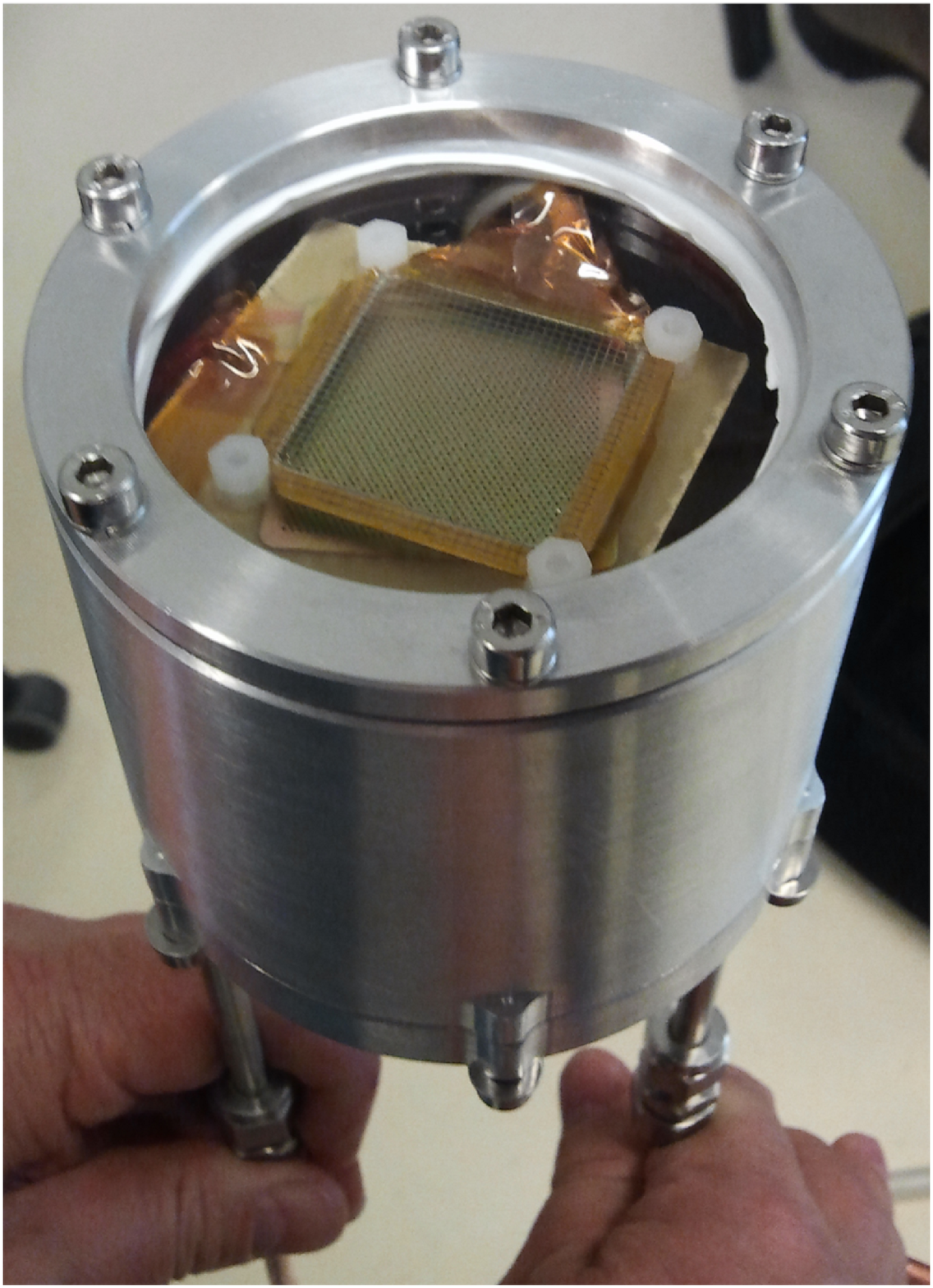}
      \caption{}
      \label{subfig:GPMstanding}
    \end{subfigure}
    \begin{subfigure}{0.45\textwidth}
      \includegraphics[width=\textwidth]{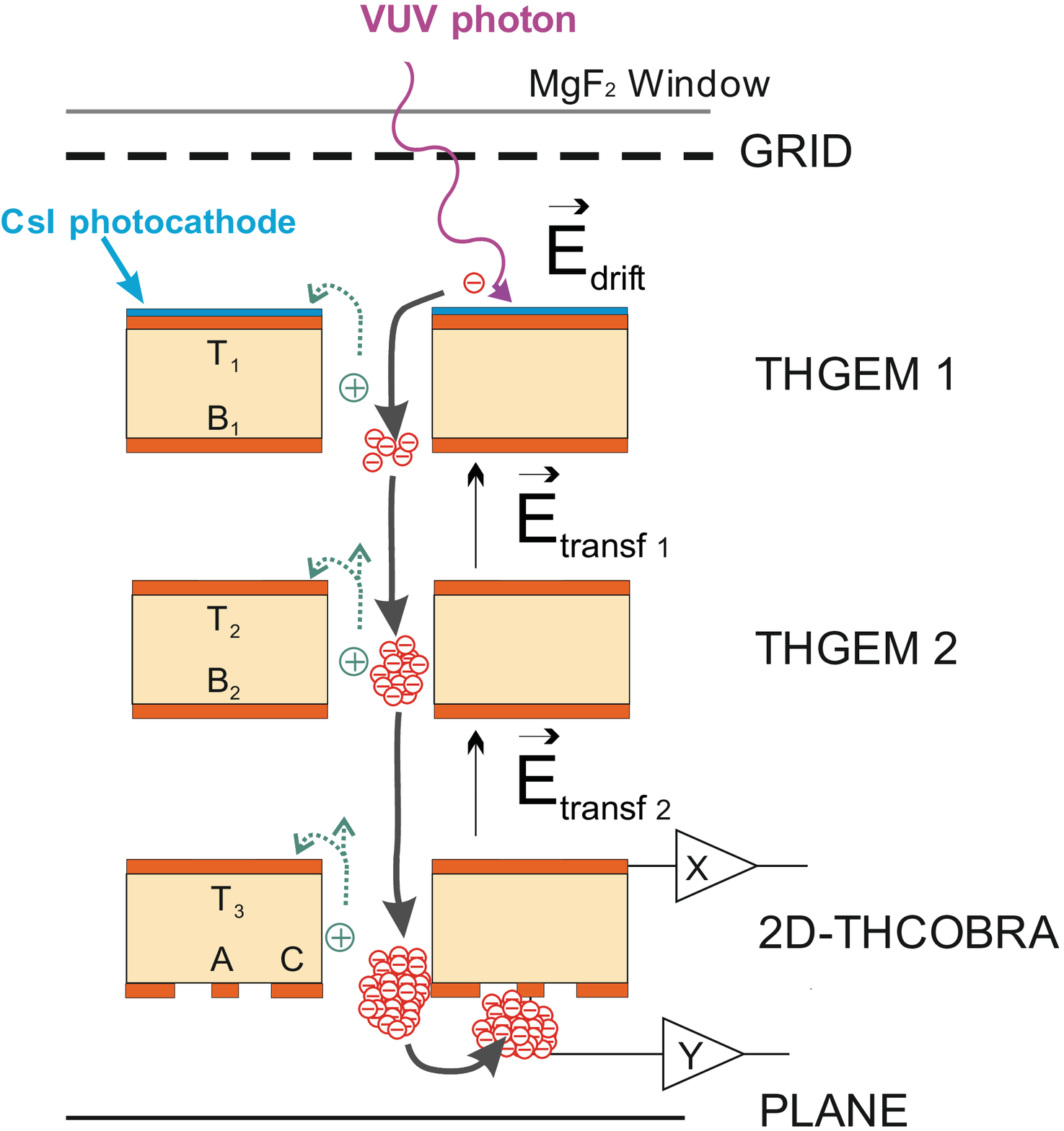}
      \caption{}
      \label{subfig:PACEM}
    \end{subfigure}
    \caption{(a) Detector prototype - the MgF$_2$ window, vacuum-sealed with Teflon gaskets to the aluminium cylinder, the grid and the first THGEM are visible, (b) schematic representation of the detector and its operation principle.}
    \label{fig:GPMov}
  \end{center}
\end{figure}

The detector used in this work is shown in Figure~\ref{subfig:GPMstanding}. The design comprises three micropatterned structures housed in an aluminium cylinder of $10$~cm diameter and $10$~cm height, with a $3$~mm thick circular window on one end and two diametrically opposed gas inlet/outlet perforations on the other. A stainless steel CF flange with nickel pins is used for signal and power feedthrough, with all structural components vacuum-sealed with Teflon gaskets.
The window material is Magnesium Fluoride (MgF$_2$) due to the high typical transmittance in the VUV range ($T>50\%$ at $121$~nm~\cite{crystranmgf2}).
The three micro-structures comprise two THGEMs and a 2D-THCOBRA, copper-cladded $400$~$\mu$m thick G10 sheets with $400$~$\mu$m holes mechanically drilled, without rim and with $800$~$\mu$m pitch in the case of the THGEMs and with an $80$~$\mu$m rim and a $1$~mm pitch in the case of the 2D-THCOBRA. Strips $100$~$\mu$m wide were etched on the top and bottom surfaces of the 2D-THCOBRA following a photolithography process, and these are joined by resistive lines deposited by serigraphy (see Figure~1 in~\cite{PositionVUV}).
The detector is operated in flow mode with a gas mixture of $95\%$ Neon and $5\%$ CH$_4$ that fills the inside of the cylinder and serves as multiplication medium. As shown in Figure~\ref{subfig:PACEM}, a VUV photon that enters through the MgF$_2$ window and interacts in the CsI photocathode deposited on top of the first THGEM may produce emission of a photoelectron with a certain probability. The photoelectron drifts due to the electric field between the THGEM top and bottom surfaces ($T_1$, $B_1$) into the THGEM holes. As it accelerates in the gas medium, collisions with Neon atoms start an electron multiplication process. The electron cloud extracted from the first THGEM holes drifts towards the second THGEM due to the transfer field $\vec{E}_{\mbox{\tiny transf 1}}$. A second multiplication occurs in the second THGEM and the electron cloud then drifts towards the 2D-THCOBRA in $\vec{E}_{\mbox{\tiny transf 2}}$. A bias voltage $V_{CT}$ is applied between the top strips ($T_3$) and the cathode ($C$) on the 
bottom of the structure, generating a field in which the electron cloud accelerates and multiplies. Further multiplication occurs due to the voltage $V_{AC}$ between the cathode and the anode strips ($A$), where the signal is collected and divided. An opposite sign signal is induced in the top strips~\cite{PositionVUV}, allowing for 2D reconstruction of the position of incidence of the VUV photon.

\section{Experimental Setup and Methods}

Detector gain measurements and image acquisition are performed simultaneously at room temperature.
For all measurements, a collimated Hg(Ar) lamp is used to provide 
the VUV photons producing the signals.
The signals from the top and anode resistive lines in the 2D-THCOBRA are preamplified 
with a Cremat CR-111 and digitized with a CAEN N1728B NIM ADC module (4 channels, 14 bits, 
$100$~MHz sampling rate) and
the image is reconstructed by weighting the 
integrated signals from each end of the resistive lines 
following the principle of resistive charge division~\cite{MHSPPosition}.

The single photoelectron energy distribution is well modelled by a Polya distribution (see e.g.~\cite{thepolya}) of the form 
\begin{equation}
\label{eq:polya}
 P_{m}(g) = \frac{m^m}{\Gamma(m)}\frac{1}{G}\left(\frac{g}{G}\right)^{m-1}e^{-m\frac{g}{G}}
\end{equation}
where $g$ is the energy, $m$ a dimensionless real parameter and $G$ the detector Gain. In log scale the function has a linear component of the form
\begin{equation}
\label{eq:linpolya}
 \log(P_{m}(g)) \sim m\frac{g}{G} + \cdots
\end{equation}
Relative gain comparisons can be performed using the inverse slope of the linear part of the distribution~\cite{2006NIMPA.558..475S, mormannthesis, Richter2002538}.

To measure the photoelectron collection efficiency of the detector, 
one end of the anode-strip resistive line is disconnected, so that all the charge flows to the other end. 
After preamplification, the signal is amplified with a Canberra 2022 (shaping time $2$~$\mu$s, $G = 100$ and $G_f = 1$) and then digitized using an Amptek MCA8000A.

\section{Results: detector characterisation}

To achieve optimal performance, the detector must hold 
its structural integrity at liquid argon temperatures and retain a stable and predictable gain. 
Due to the nature of the liquid argon scintillation light, the detector has to detect 
simultaneous multiple photon interactions. In this section, measurements to determine the detector performance are presented.
First, measurements to characterise the GPM behaviour are performed at room temperature: gain, photoelectron collection efficiency and position resolution, and finally, preliminary tests to evaluate the detector behaviour under simulated liquid argon conditions are carried out: multiphoton position reconstruction, gain stability and evolution and detector structural integrity tests at cryogenic temperatures.

The detector gain was measured at room temperature as a function of the two 2D-THCOBRA potentials, $V_{AC}$ and $V_{CT}$, with the potentials on THGEM1 and THGEM2 fixed at $595$~V and $550$~V respectively, empirically chosen as a compromise between efficiency and discharge probability, and the transfer fields set to $E_{\mbox{transf 1}} = E_{\mbox{transf 2}} = 300$~V/cm.
\begin{figure}[htbp]
  \begin{center}
    \begin{subfigure}{0.45\textwidth}
      \includegraphics[width=\textwidth]{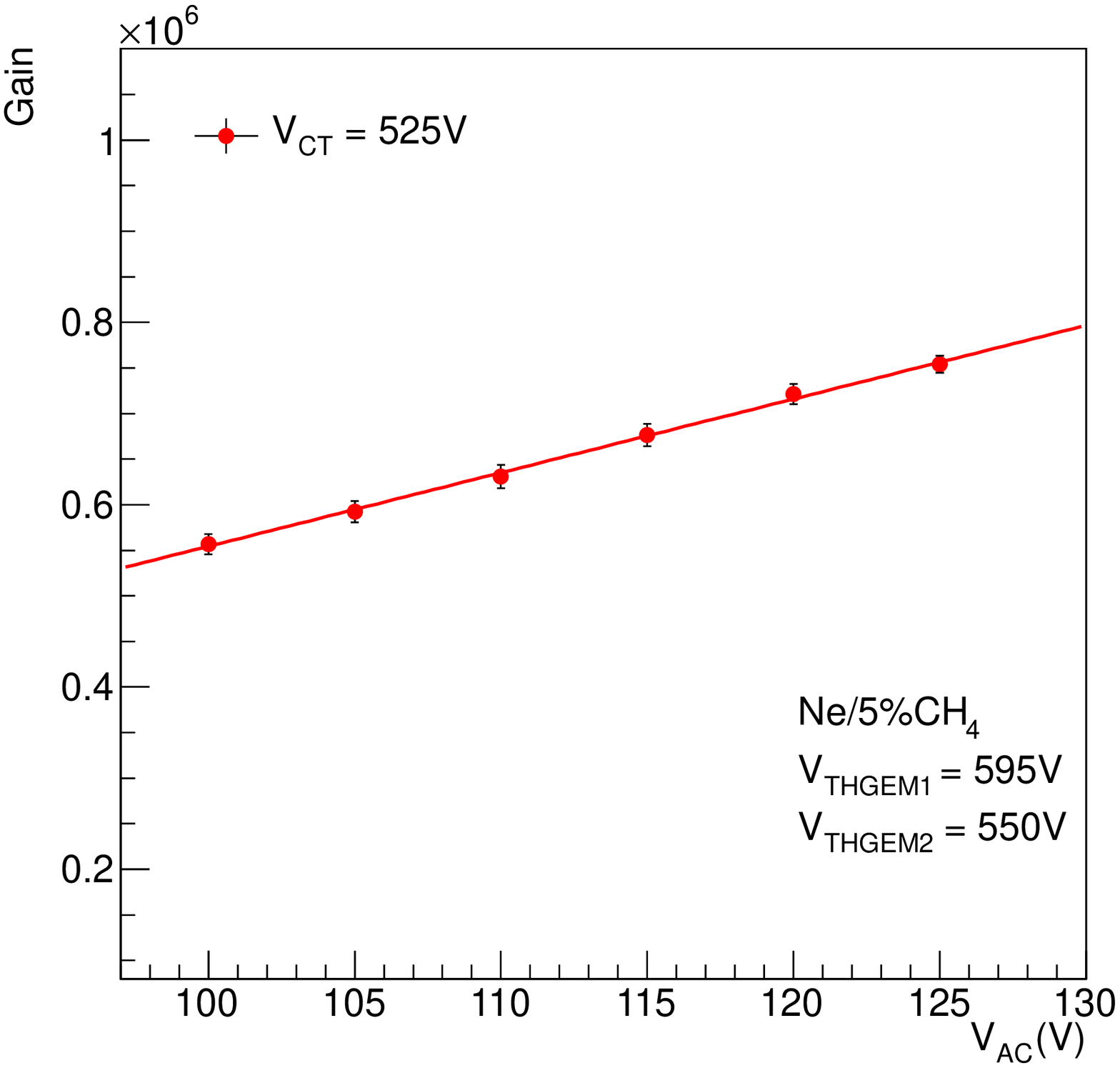}
      \caption{}
      \label{fig:VACGain}
    \end{subfigure}
    \begin{subfigure}{0.45\textwidth}
      \includegraphics[width=\textwidth]{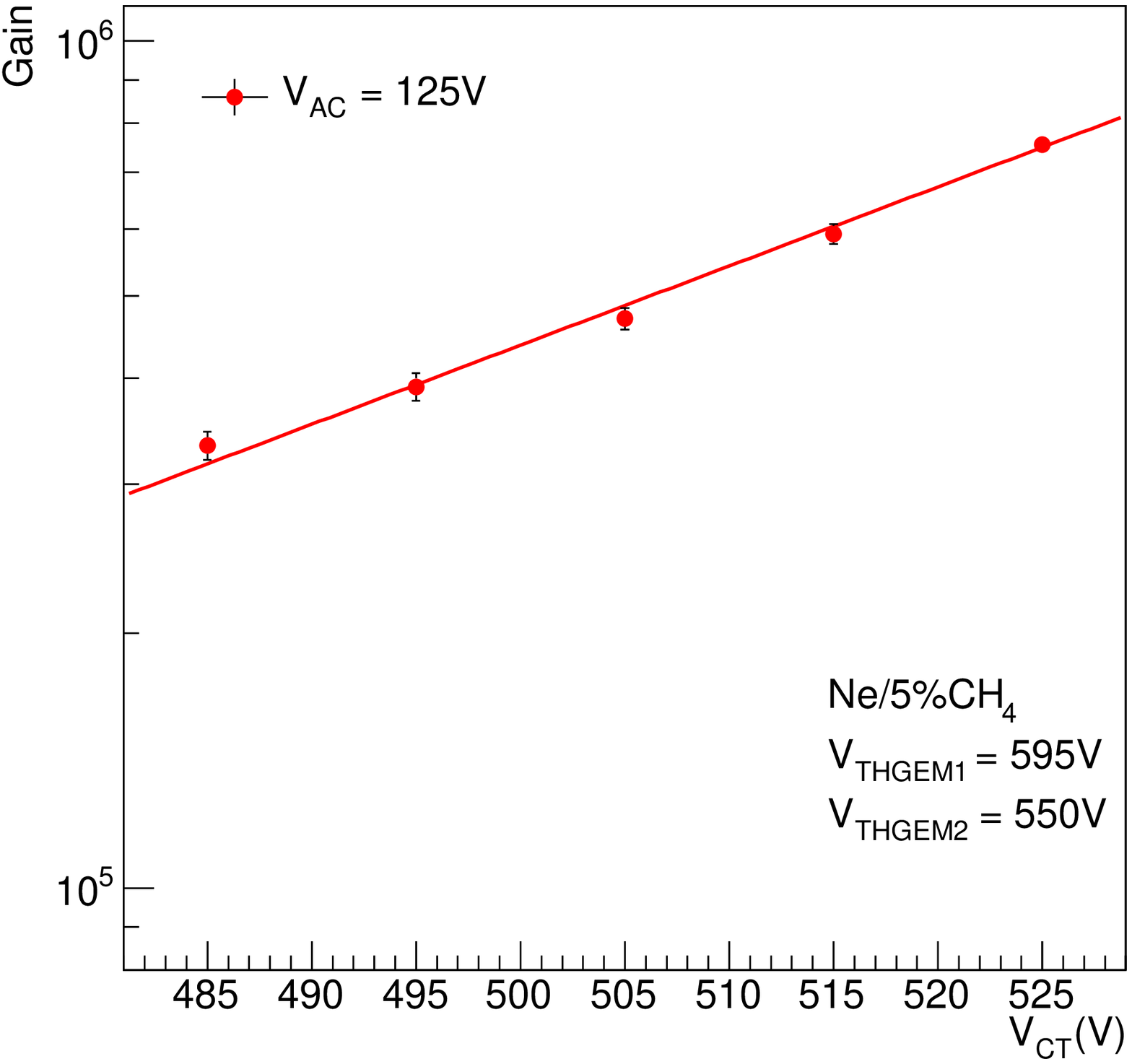}
      \caption{}
      \label{fig:VCTGain}
    \end{subfigure}
    \caption{Detector gain as a function of the Anode strips$-$Cathodes potential (a), and 
as a function of Top strips$-$Cathodes potential (b), in number of collected electrons per photoelectron.}
    \label{fig:Gains}
  \end{center}
\end{figure}
The gain as a function of the Anode strips$-$Cathode potential $V_{\mbox{AC}}$ is shown in Figure~\ref{fig:VACGain} 
for $V_{\mbox{AC}}$ from $100$~V to $125$~V in steps of $5$~V and for Cathodes$-$Top 
strips potential fixed at $V_{\mbox{CT}} = 525$~V. 
The behaviour observed in this region is approximately linear, 
indicating that no additional electron multiplication occurs in the gas at this stage. 
In Figure~\ref{fig:VCTGain}, gain measurements as a function of $V_{\mbox{CT}}$ are presented. 
$V_{\mbox{CT}}$ is varied from $485$~V to $525$~V in steps of $10$~V for a constant $V_{\mbox{AC}} = 125$~V. 
In this range, the observed variation has an exponential behaviour, as there is electron multiplication taking 
place in the 2D-THCOBRA holes, between the top strips and the cathodes. Based on this study, optimal operation voltages were chosen ($V_{\mbox{AC}} = 125$~V and $V_{\mbox{CT}} = 525$~V), corresponding to a gain of $G = 8\cdot10^{5}$. For these values, detector operation is stable and there is a low discharge rate.

\begin{figure}[htbp]
  \begin{center}
    \includegraphics[width=0.5\textwidth]{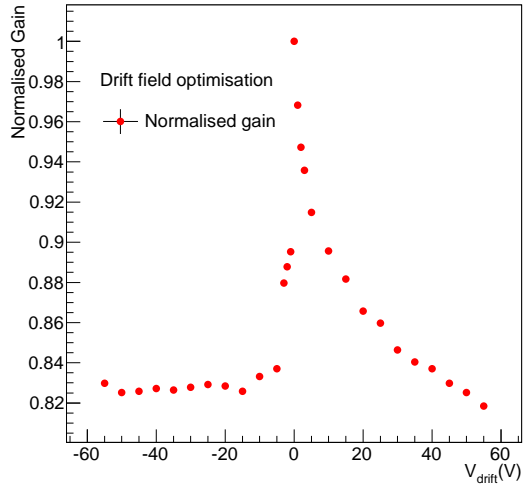}
    \caption{Detector gain as a function of drift potential}
    \label{fig:Drift}
  \end{center}
\end{figure}

The drift field between the first THGEM and the grid has
a strong effect on the extraction of photoelectrons from the CsI surface.
In Figure~\ref{fig:Drift}, the detector gain is plotted as a function 
of the drift potential with respect to the top of THGEM1, where
it is shown that the gain falls rapidly when applying a non-zero potential.
A negative potential stops the photoelectrons 
from being extracted from the CsI layer and the gain flattens out at $\sim83\%$ 
of the maximum for $V_{\mbox{drift}}\simeq -10$~V.
However, with increasingly positive potential the extraction of photoelectrons is aided by the additional electric field, and the gain drops more slowly. 
The optimal value found was $V_{\mbox{drift}} = (0\pm0.5)$~V, so the grid potential 
was set to zero with respect to the first THGEM for the rest of the tests.

\begin{figure}[htbp]
  \begin{center}
    \includegraphics[width=0.5\textwidth]{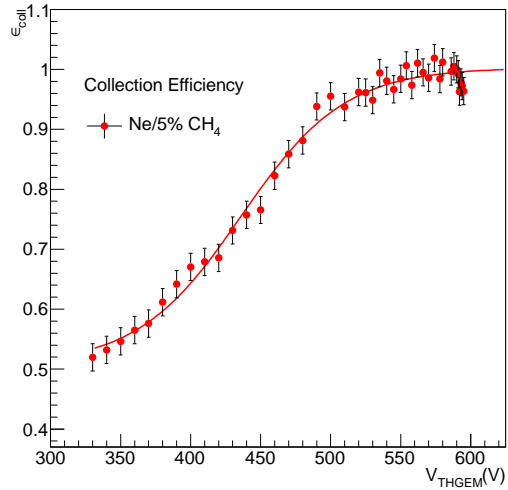}
    \caption{Photoelectron collection efficiency as a function of THGEM1 potential}
    \label{fig:ColEff}
  \end{center}
\end{figure}

Maximising the detector collection efficiency (ratio of collected to extracted photoelectrons) is specially important when working in single-photoelectron mode. This ratio approaches 1 as $V_{\mbox{THGEM1}}$ is increased, and to measure it the gain is kept approximately constant for different voltage settings by comparing a 
reasonably linear region of the energy spectra and applying equation~\ref{eq:linpolya}. 
The comparison is performed by integrating this region to estimate the amount of collected charge for different THGEM1 potentials~\cite{CollExtractionEff, Veloso2011}. 
The result is shown in Fig~\ref{fig:ColEff}, where $\epsilon_{\mbox{coll}}\sim1$ at $V_{\mbox{THGEM1}} = 595$~V.
At this bias voltage, the surface field between holes is high enough for the detector to reach an extraction efficiency higher than $70\%$~\cite{CollExtractionEff}.

\begin{figure}[htbp]
  \begin{center}
    \includegraphics[width=0.9\textwidth]{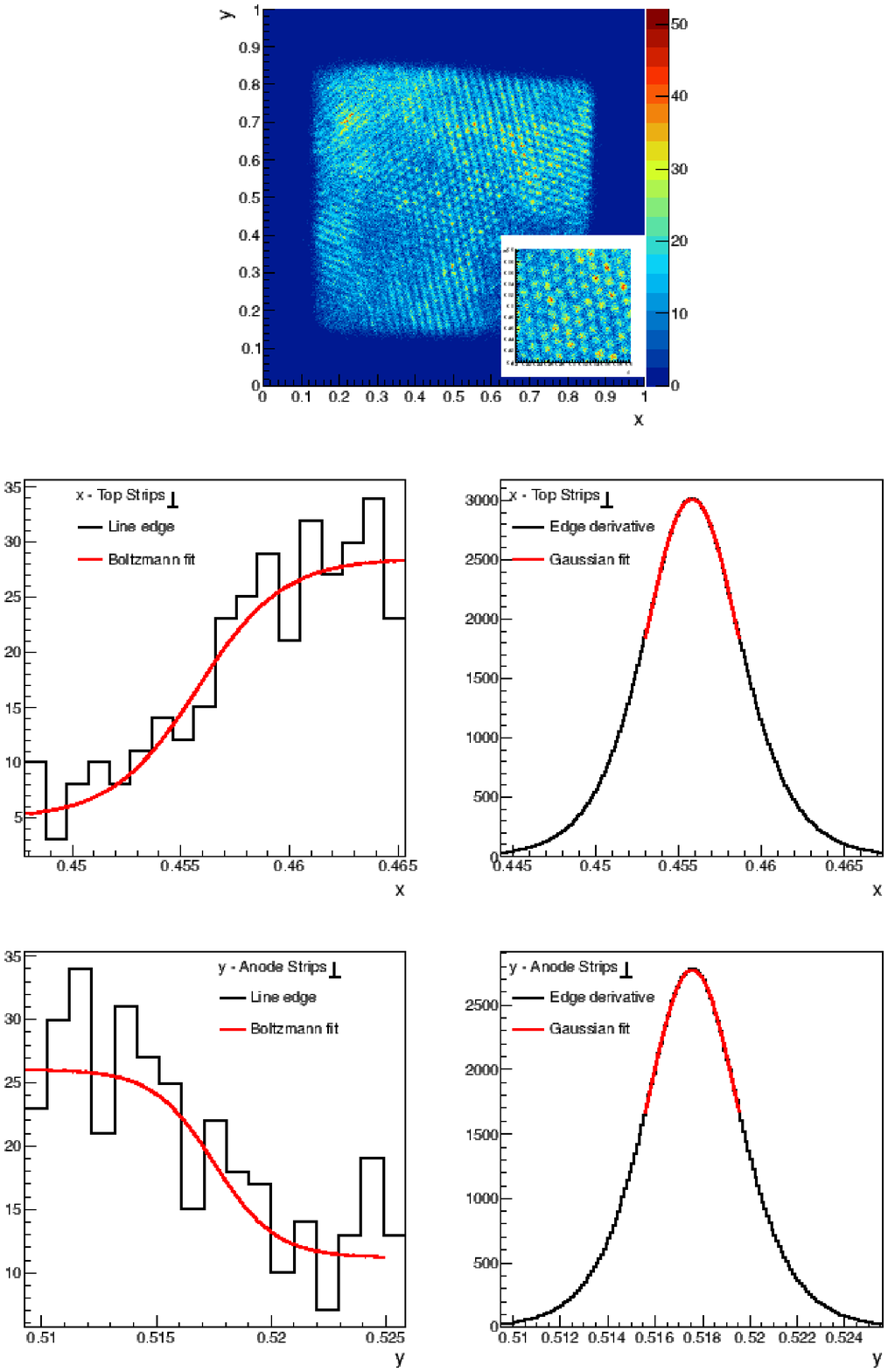}
    \caption{Position resolution measurement with the Line Spread Function method. The spread of one of the edges in the THGEM1 imaged pattern is fitted to a Boltzmann function. The derivative of the fit is taken and fitted to a gaussian to obtain the resolution. The results obtained are $\sigma = 90\pm30$~$\mu$m in the $x$-direction (top strips) and $\sigma \lesssim 90\pm30$~$\mu$m in the $y$-direction (anode strips). Scale: $3.12$~cm $= 1$.}
    \label{fig:PosRes}
  \end{center}
\end{figure}

To determine the GPM position resolution the edge spread function method was applied 
to one of the edges of the pattern created by the first THGEM in the image (see Figure~\ref{fig:PosRes}). 
The result of the fit yields $<90\pm30$~$\mu$m in the direction of the anode strips 
and $90\pm30$~$\mu$m in the direction of the top strips.

\begin{figure}[htbp]
  \begin{center}
    \begin{subfigure}{0.45\textwidth}
      \includegraphics[width=\textwidth]{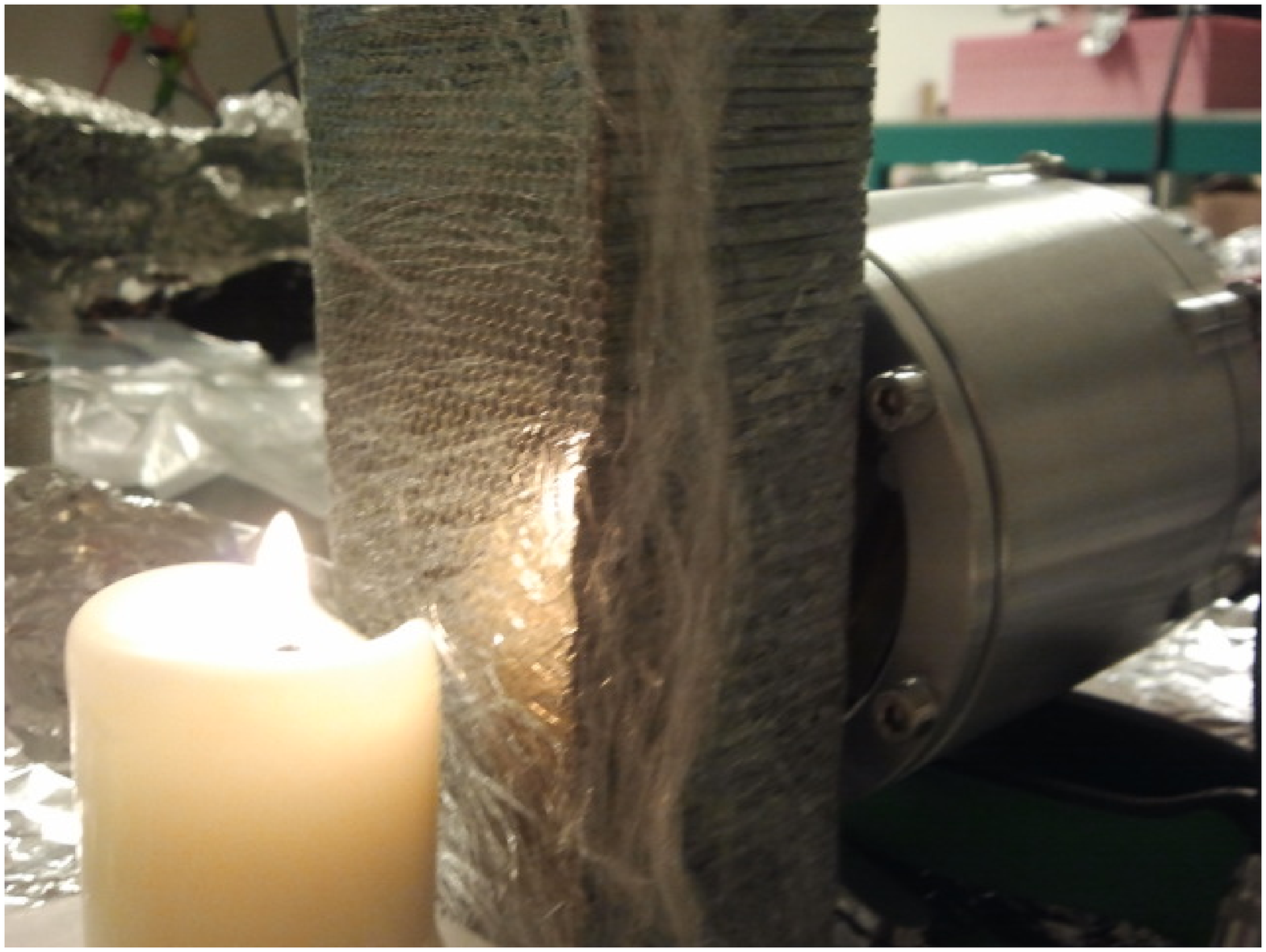}
    \end{subfigure}
    \begin{subfigure}{0.45\textwidth}
      \includegraphics[width=\textwidth]{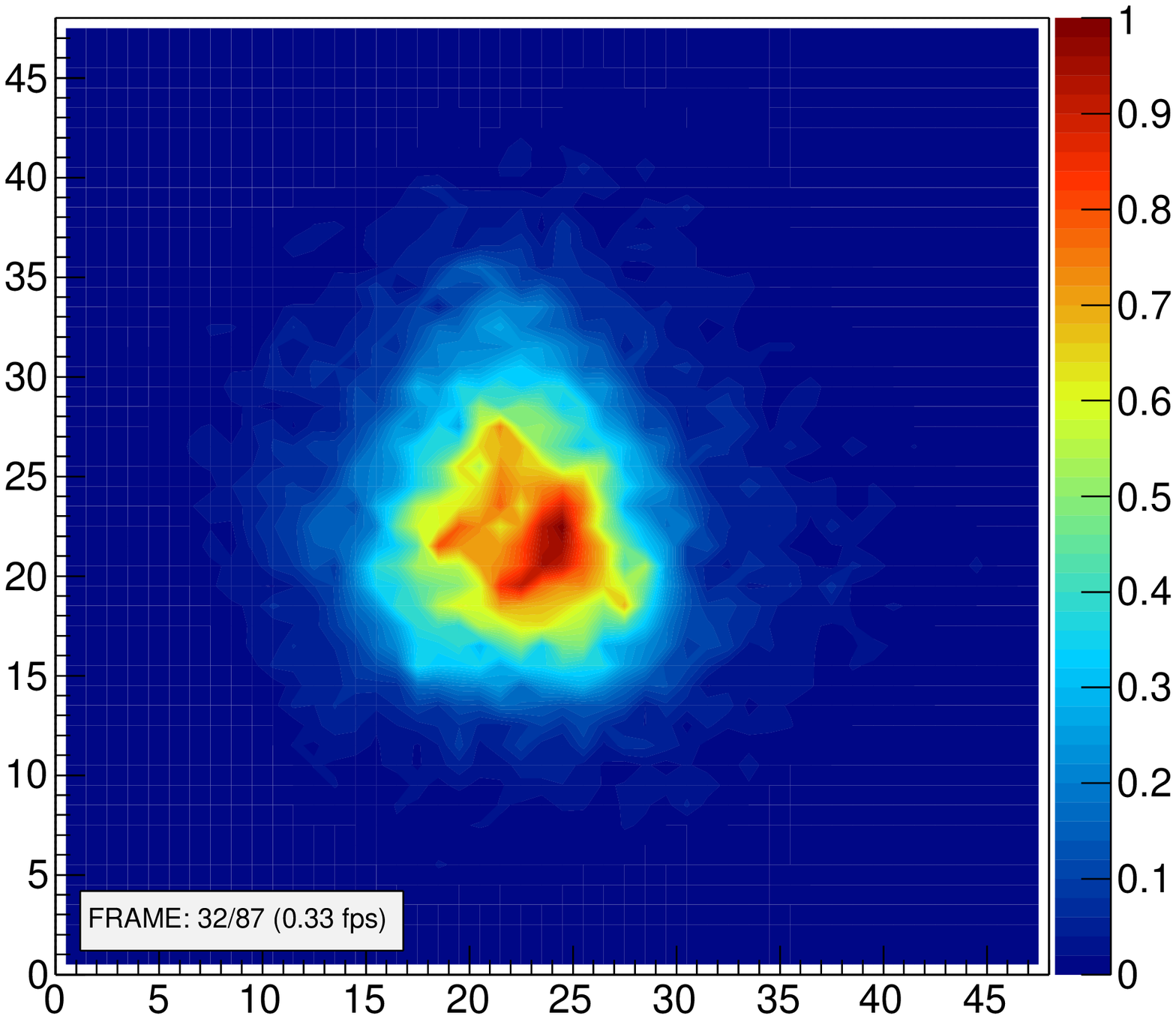}
    \end{subfigure}
    \caption{Experimental setup for candle
      tests. A lit candle in front of the GPM, UV radiation
      collimated and attenuated with plastic
      film (left). Example image obtained from a
      $3$~s exposure time, showing the burning candle flame shape (right).
      A full animated sequence can be watched in~\cite{candlemovie}. Scale: $3.12$~cm $= 48px$.}
    \label{fig:Candle}
  \end{center}
\end{figure}

An experiment was set up to test the detector ability to reconstruct UV light from naturally occurring sources. Flame light below the CsI $220$~nm threshold should be detectable by the GPM~\cite{PSGPMreview2008}, so a lit candle was placed in front of the detector, collimated and attenuated with plastic film, as shown in Figure~\ref{fig:Candle}~(left). In the absence of attenuation, flame UV light overwhelmed the detector, causing discharges due to space-charge accumulation, confirming the hypothesis. With enough attenuation, it was possible to ensure that only single photons hit the detector. An image built after a $3$~s exposure to candle light is shown in Figure~\ref{fig:Candle}~(right). A series of frames were recorded and a movie showing the movement of the UV light sources within the flame can be found in~\cite{candlemovie}. In combination with an IR detector, a $360^{\circ}$ collimated GPM can be used for outdoor fire detection.

Liquid argon produces $<51000$ scintillation photons per MeV of incident particle~\cite{LARPHOTONSPERKEV}.
These photons can extract photoelectrons from the GPM CsI photocathode ($QE > 10\%$~\cite{CsIQE}). Assuming that the detector is observing
$0.511$~MeV gamma-ray interactions in a cylinder of liquid argon of $10$~cm height with a MgF$_2$ window ($70$~mm diameter),
the number of photoelectrons generated is on the order of $\mathcal{O}(10^1)\sim\mathcal{O}(10^2)$.
Therefore, the system must be able to reconstruct events with multiple photons 
that simultaneously produce photoelectrons in the photocathode.

\begin{figure}[htbp]
  \begin{center}
    \includegraphics[width=1.0\textwidth]{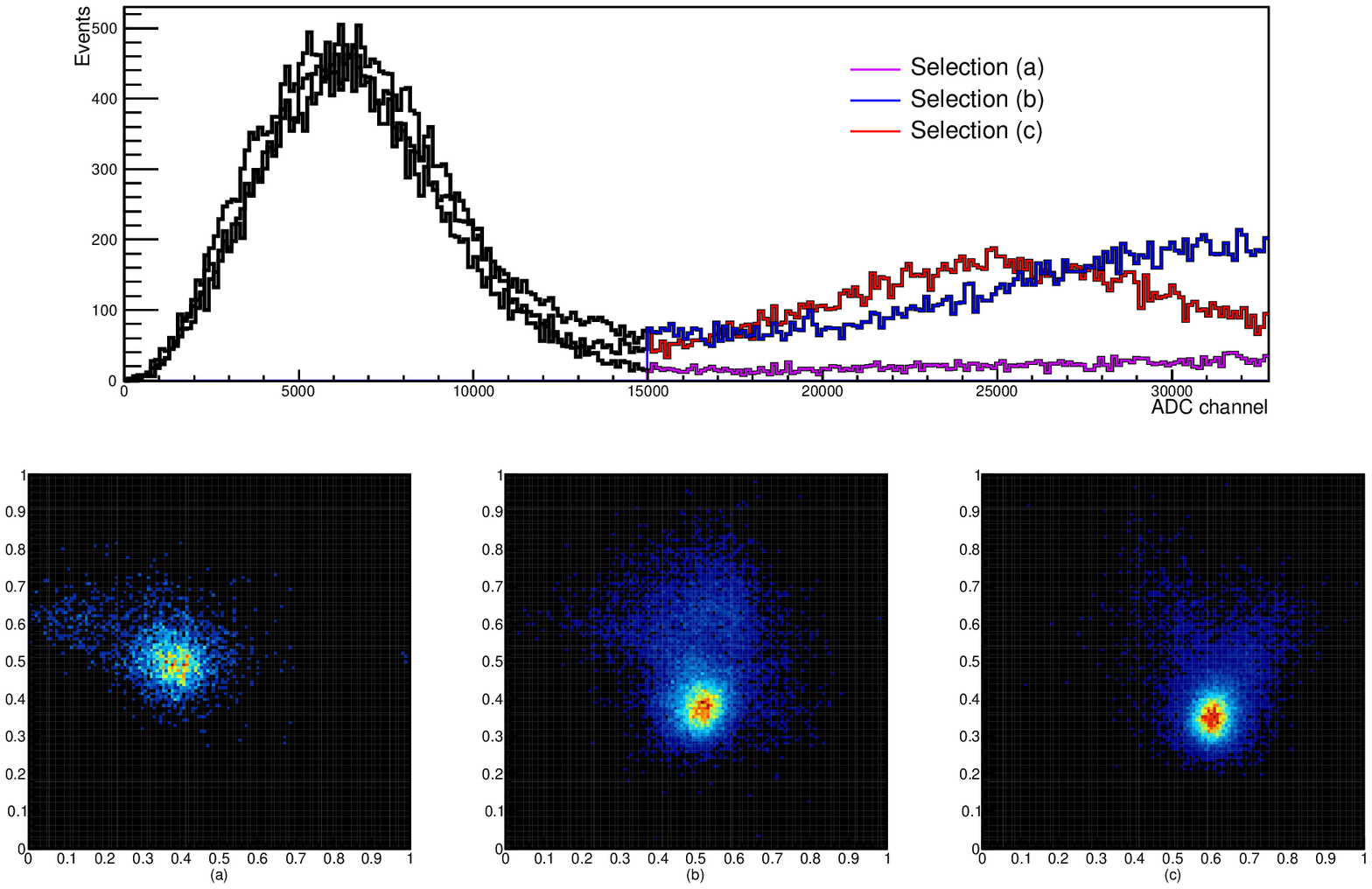}
    \caption{Displacement of average position and energy distribution of multiple photon interactions. The single-photon polya energy distribution is modified by multiple photon interactions appearing in the high end of the spectrum as a resonant peak, revealing the average energy deposited in the detector per multiphoton event. A different average number of simultaneous photons interact in the photocathode in cases (b) and (c), and hence their energy distributions do not peak at the same energy. The maximum number of available ADC channels was reached and the distributions cannot be fully shown. An excessive number of low-energy events during data-taking hides the multiphoton peak in (a) due to pileup. Scale: $3.12$~cm $= 1$.}
    \label{fig:MPI}
  \end{center}
\end{figure}

To simulate argon emission, a spinning disk with a slit allows pulses of 
UV light from the Hg(Ar) lamp to be detected, being integrated for $10$~$\mu$s.
The detector, operated at room temperature at $20$~cm from the light source was moved from right to left in steps of $\sim2$~cm
to determine if the average position of the photon 
pulses would vary accordingly. In Figure~\ref{fig:MPI} three energy distributions and their corresponding position distributions are shown. 
The results show that the detector can simultaneously reconstruct the energy of multiple 
photons and their average position of arrival at the photocathode.
The first measurement (violet energy distribution, position (a) ) suffered from 
more noise from one of the channels, hiding most of the multiple photon signal due to pileup conditions. 
The trigger threshold was increased for the rest of the measurements.
It is worth noting that when operated in multiphoton mode, the detector loses its single photon position capability
(reconstructed always around the centre)
 due to the long integration time.

In order to test the stability of the gain during prolonged time periods, the GPM was set up to detect 
single photons from the UV Hg(Ar) lamp. The experiment ran 
for approximately $44$~h, collecting data in intervals of $3$~min.
The following variables were measured: anode and top-strips gain, pressure of the gas entering and leaving the detector, room temperature and instantaneous discharge times. The purely exponential part of the measured energy distributions for every $3$~min interval was fitted to extract the gain. Every signal indicating a high current in the voltage supply (discharge) was recorded.

\begin{figure}[htbp]
  \begin{center}
    \includegraphics[width=0.5\textwidth]{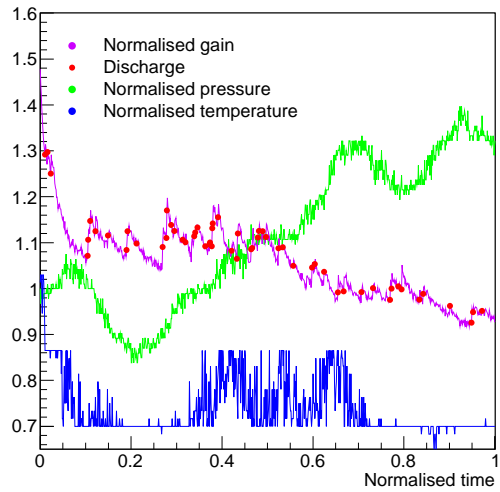}
    \caption{Normalised pressure (green histogram), temperature (blue histogram) and gain from 
the anode strips channels (purple histogram), as a function of time. 
Discharges are shown as red points on the gain distribution.
}
    \label{fig:VarsvsTime}
  \end{center}
\end{figure}

\begin{figure}[htbp]
  \begin{center}
    \includegraphics[width=0.5\textwidth]{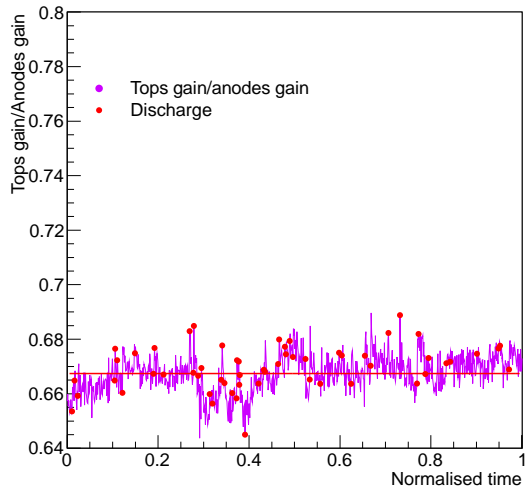}
     \caption{Ratio of the gain measured from the top strips signal to the gain measured from the anode strips signal. A stable value of $\sim67\%$ is observed during operation.}
    \label{fig:G3/G1}
  \end{center}
\end{figure}

A standard Principal Component Analysis (PCA) reveals linear dependence between gains, 
room temperature and pressure of the gas leaving the detector. In Figure~\ref{fig:VarsvsTime}, 
the normalised pressure (green histogram) and temperature (blue histogram) are shown together with the gain from 
the anode strips channels (purple histogram), as a function of time\footnote{Gain is normalised to an 
arbitrary central value ($1200$~ADC$/e^-$). Pressure is normalised to the first measured value
($1009.22$~hPa) and temperature to the most common value ($297.1$~K). 
Pressure and temperature are then transformed as $1 + 100\times (x - 1)$, and temperature 
is shifted down by 0.3 for illustration purposes. Time is normalised to the length of the experiment 
($145711$~s or $40.48$~h)}. Discharges are shown as red points on the gain distribution. The ratio between top and anode strips gain is shown in Figure~\ref{fig:G3/G1}, stable at a value of~$\sim67\%$.

\begin{figure}[htbp]
  \begin{center}
    \includegraphics[width=0.5\textwidth]{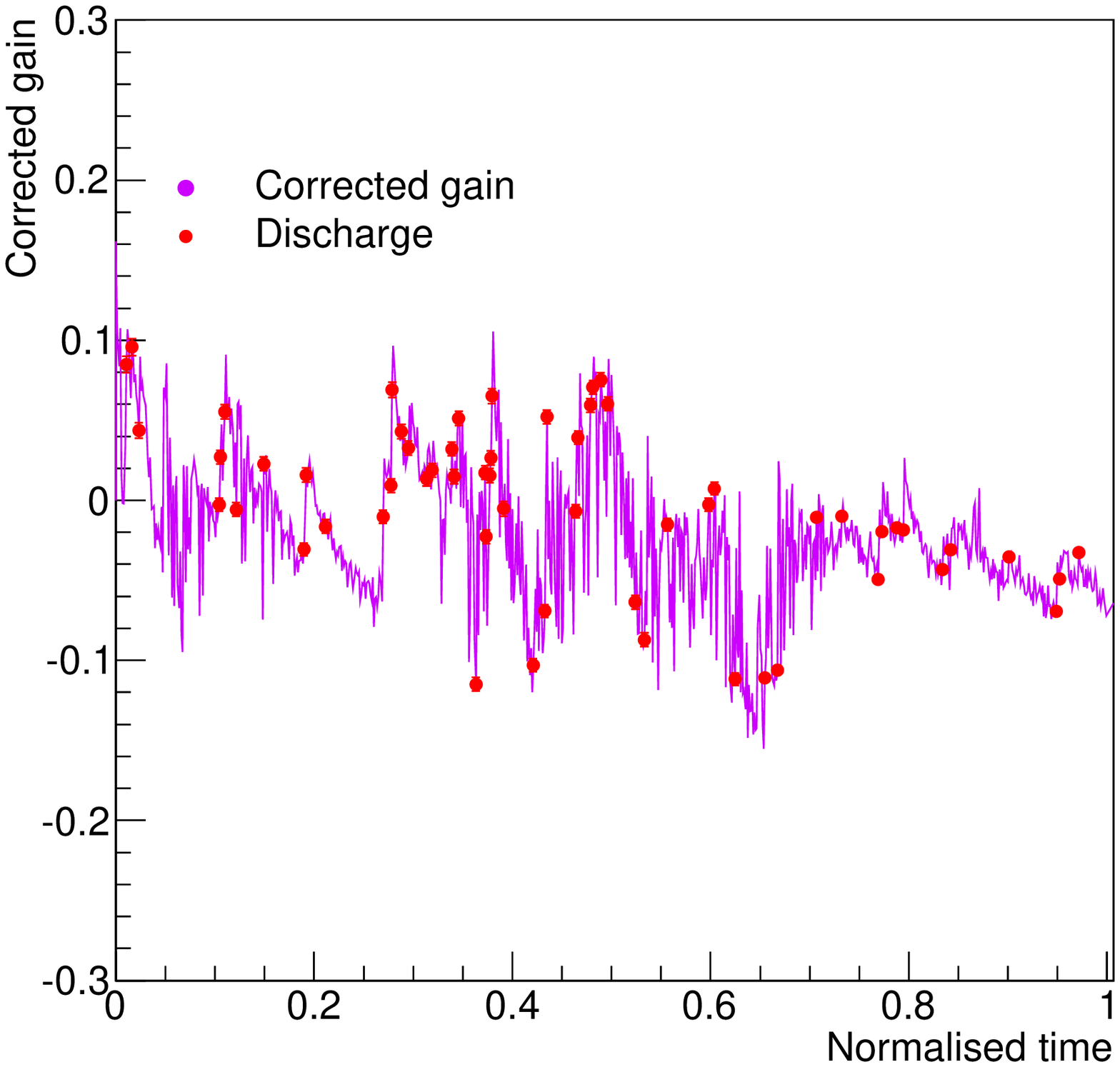}
    \caption{Gain evolution after correcting for the variation due to temperature and pressure. Discharges induce oscillations of $\pm10\%$ around the mean.}
    \label{fig:corrG1}
  \end{center}
\end{figure}

The variation in gain observed in Figure~\ref{fig:VarsvsTime} is mainly due to the variation in room 
temperature and pressure and to occasional discharges, and can be described by:
\begin{equation}
 \frac{dG}{dt} = \frac{\partial G}{\partial p}\frac{dp}{dt} + \frac{\partial G}{\partial T}\frac{dT}{dt} + \left.\frac{\partial G}{\partial t}\right\rfloor_{\mbox{\tiny discharge}}
\end{equation}
Where the partial derivatives are calculated for all the other variables constant. Integrating this expression yields:

\begin{equation} \label{eq1}
\begin{split}
G(p(t), T(t), \mbox{\small disch}(t)) & = G_0 + \frac{\partial G}{\partial p}p(t) + \frac{\partial G}{\partial T}T(t) + G(\mbox{\small disch}(t)) \\
 & = G_1(p(t)) + G_2(T(t)) + G_3(\mbox{\small disch}(t))
\end{split}
\end{equation}
The behaviour of the gain with respect to pressure variations $G_1(p)$ was studied in regions 
of constant room temperature ($T = 297.1$~K). On average, a discharge occurred every $46$~minutes, 
so it was required that no discharge occurred in the last~$30$~mins.
 A linear correlation between pressure and gain was established, with negative slope. 
Analogously, $G_2(T)$ was fitted and a positive slope was found.
The discharge-dependent term is obtained by calculating~$G_3(\mbox{\small disch}(t)) = G(p, T, disch) -  G_1(p) - G_2(T)$. As shown in Figure~\ref{fig:corrG1}, 
discharges are responsible for a~$\pm10\%$ variation around the mean at~$(0\pm2)\%$.

Given the stable operation and multiple photon detection capabilities, preliminary structural tests at 
cryogenic temperatures were carried out. Initially, the MgF$_2$ window was substituted for a more robust
aluminium window. The detector was then evacuated to a pressure of $\sim10^{-6}$~mbar and then cooled down to $77$~K with liquid Nitrogen. 
After reaching stability, the detector was removed from the liquid and was left to heat up to room temperature.
During this stage, the pressure increased to $200$~mbar, an encouraging result
considering that the gas pressure will be kept at $\sim1$~atm during normal operation. 
To further ensure structural integrity and good performance of the Teflon gaskets, an additional test was performed with a dummy glass window.
argon was allowed to flow through the detector until all air had been removed. 
When the pressure reached $1.2$~atm, all valves were shut and the detector 
was cooled down using liquid Nitrogen.
The pressure was maintained at $\sim1.2$~atm by adding more argon, since the gas freezes below $84$~K. 
When equilibrium was reached, the liquid Nitrogen was removed and a block of 
solid argon was visible through the glass window of the detector. As the system warmed up to room temperature, 
the flow of escaping argon was regulated to keep the pressure constant.
When the pressure dropped to $1$~atm all valves were shut.
No water, in liquid or solid state, was visible inside the detector volume or on the inner surface 
of the glass window after all the argon evaporated and the system reached room temperature.
The detector was pumped down to $10^{-6}$~mbar to test the glass window strength, without problems. 
After turning off the pump the pressure did not go above $10$~mbar, hence the system is vacuum tight
with Teflon gaskets and a glass window, even after the process of cooling and heating. 
While in the actual prototype a MgF$_2$ window is used, there is no reason to expect a significant 
change with respect to the measurements performed with glass.

\section{Conclusions}

A new large area  Gaseous Photomultiplier utilizing a cascade of Thick GEM layers 
intended for gamma-ray position reconstruction in liquid argon is proposed.
A prototype designed to operate at cryogenic temperatures 
inside the liquid phase and to reconstruct liquid argon scintillation light 
was built. A number of performance measurements were carried out at room temperature and stable operation high-voltage settings: 
photoelectron collection efficiency, position resolution and 
stability.
A photoelectron collection efficiency on the order of~$1$, a gain of $8\cdot10^{5}$ per photoelectron 
and a position resolution better than  100$~\mu$m
were measured.
Discharges were observed every $46$~min
of operation on average ($0.1$~discharges/($h\cdot cm^2$)).
Detector gains were stable for the whole data taking
period within $\pm10\%$, showing a
slow charging-up effect after every discharge.
Gain variations due to pressure and temperature disappear when the two
variables are under control in the laboratory. However, to reduce the
discharge rate it is necessary to operate with lower bias settings,
limiting the stability of the gain to the above-mentioned $\pm10\%$.

The proposed detector has potential applications ranging from medical physics and engineering, to particle physics.
Initial tests of robustness against cryogenic temperatures were performed successfully.
The next essential step would be to demonstrate the operation of the Gaseous Photomultiplier inside the liquid argon phase.

\acknowledgments
Brais Lopez Paredes was supported by a scholarship from the University of Sheffield Physics and Astronomy Department.
Carlos Azevedo was supported by PostDoctoral grant from FCT (Lisbon) SFRH/BPD/79163/2011.
Ana Luisa Silva was supported by QREN programme Mais Centro, FEDER and COMPETE, through project CENTRO-07-ST24-FEDER-002030. 
This work took place in the i3N DRIM laboratory at the University of Aveiro. We acknowledge support from the University of Sheffield
PoC and MRC award numbers X/008226 and R/139662.

\bibliography{references}

\end{document}